\documentclass[9pt, conference]{IEEEtran}
\usepackage{waspaa25} % removes hyperlinks (required by IEEE Xplore)

\usepackage{acronym}
\usepackage{bm}
\usepackage{amsmath}
\usepackage{enumitem}
\usepackage{pbalance}

\graphicspath{{./figures/}}

\acrodef{BWE}{bandwidth extension}
\acrodef{UBGAN}{Universal Bandwidth Extension Generative Adversarial Network}
\acrodef{WB}{wideband}
\acrodef{SWB}{super-wideband}
\acrodef{AMR-NB}{Adaptive Multi-Rate narrowband}
\acrodef{EVS}{Enhanced Voice Services}
\acrodef{BBWE}{blind-BWE}
\acrodef{CELP}{Code-Excited Linear Prediction}
\acrodef{TBE}{time domain bandwidth extension}
\acrodef{MFCCs}{Mel Frequency Cepstral Coefficients}
\acrodef{FB}{Fullband}
\acrodef{STFT}{short-time Fourier transform}
\acrodef{PQMF}{Pseudo-Quadrature Mirror Filters}
\acrodef{GBWE}{guided-BWE}
\acrodef{DPCRNN}{Dual-Path Convolutional Recurrent Network}
\acrodef{GRU}{Gated Recurrent Unit}
\acrodef{SQ}{scalar quantization}
\acrodef{LP}{Linear Prediction}
\acrodef{}{}
\acrodef{}{}

\title{UBGAN: Enhancing Coded Speech with Blind and Guided \\Bandwidth Extension}
\name{Kishan Gupta$^{1}$,
      Srikanth Korse$^{2}$,
      Andreas Brendel$^{1}$,
      Nicola Pia$^{1}$,
      Guillaume Fuchs$^{1,2}$}
\address{$^{1}$Fraunhofer IIS, Erlangen, Germany\\ 
$^{2}$International Audio Laboratories Erlangen, Germany.\\
}

\begin{document}
\maketitle

%To enhance the perceptual quality of coded speech without compromising other aspects, \ac{BWE} of the transmitted speech is an attractive and popular technique in conventional coding. However, neural speech codecs often lack modularity, as they are often designed to perform end-to-end coding at single fixed sampling rate without an additional module for \ac{BWE}. In this paper, we propose

\begin{abstract}
	%Most conventional and neural speech codecs operate on wideband speech signals to achieve a good compromise between achievable perceptual quality, engendered bitrate and computational complexity.

 In practical application of speech codecs, a multitude of factors such as the quality of the radio connection, limiting hardware or required user experience necessitate trade-offs between achievable perceptual quality, engendered bitrate and computational complexity. Most conventional and neural speech codecs operate on \ac{WB} speech signals to achieve this compromise. To further enhance the perceptual quality of coded speech, \ac{BWE} of the transmitted speech is an attractive and popular technique in conventional speech coding. In contrast, neural speech codecs are typically trained end-to-end to a specific set of requirements and are often not easily adaptable. In particular, they are typically trained to operate at a single fixed sampling rate. With the \ac{UBGAN}, we propose a modular and  lightweight GAN-based solution that increases the operational flexibility of a wide range of conventional and neural codecs. Our model operates in the subband domain and extends the bandwidth of \ac{WB} signals from 8~kHz to 16~kHz, resulting in \ac{SWB} signals. We further introduce two variants, guided-UBGAN and blind-UBGAN, where the guided version transmits quantized learned representation as a side information at a very low bitrate additional to the bitrate of the codec, while blind-\ac{BWE} operates without such side-information. Our subjective assessments demonstrate the advantage of \ac{UBGAN} applied to \ac{WB} codecs and highlight the generalization capacity of our proposed method across multiple codecs and bitrates.
 %a lightweight GAN-based \ac{BWE} approach specifically designed to work with various conventional and neural codecs, we propose a modular solution that increases the flexibility of a broad range of speech coding systems. 
	
\end{abstract}
\acresetall

\section{Introduction}
\label{sec:intro}

Speech codecs for real-time communications have evolved significantly over the years. With the increasing demand for higher speech quality, the limitations of traditional codecs, such as \ac{AMR-NB}~\cite{amr_nb}, became apparent due to their limited bandwidth. This led to the adoption of more advanced codecs, including AMR-WB~\cite{amr_nb}, a \ac{WB} codec with sampling rate of 16~kHz and \ac{EVS}~\cite{evs_overview}, which supports multiple bandwidths, including a \ac{SWB} mode with a sampling rate of 32~kHz. A key aspect of this evolution has been the adoption of \ac{BWE}, a technique that generates high-frequency content of speech signals from the received coded low-frequency components and possibly from other transmitted parameters. This approach is widely used alongside core codecs to enhance the quality of the decoded signals. \ac{BWE} enables codecs to allocate bits more effectively and produces more intelligible speech through the reconstruction of high-frequency content. %while minimizing quality degradation in the lower band.

\ac{BWE} has become a pivotal part of modern classical speech codecs. It is usually employed as \ac{BBWE} for the lowest bitrates or as \ac{GBWE} utilizing side-information when the bit-budget is higher. AMR-WB features a \ac{CELP} core that operates in the 0-6.4~kHz band and applies BBWE to extend the bandwidth up to 7~kHz~\cite{amrwb} by shaping a white noise signal with a interpolated \ac{LP} synthesis filter. More advanced codecs like EVS offer a CELP core at two different internal sampling rates: 12.8~kHz and 16~kHz depending on the bitrates. EVS primary mode employs a more sophisticated technique, known as \ac{TBE}~\cite{evs_swb} to obtain either WB or SWB signals. In TBE, the high frequency fine structure~(excitation) is generated by applying a non-linear operation on the low-band excitation before being spectrally shaped with an LP synthesis filter that is sent as side information along with energy gains. 

While classical codecs are still widely used, neural speech codecs have been gaining momentum in recent years as they are capable of generating high quality speech at very low bitrates. The earliest approaches~\cite{wavenet}\cite{ssmgan}\cite{lpcnet} typically focused on decoding speech using signal features such as \ac{MFCCs}, Mel-spectrograms or other acoustic parameters, and were limited to generating \ac{WB} signals due to the low resolution of these representations at higher frequency bands. The rise of end-to-end trained neural codecs improved achievable quality and enabled the support of a wide range of bandwidth such as \ac{WB}\cite{tfnet}\cite{nesc}, semi-\ac{SWB}~(24~kHz sampling rate)\cite{soundstream}\cite{encodec}\cite{dac} and even \ac{FB}~(48~kHz sampling rate)\cite{audiodec}. However, most of these end-to-end systems rely on time-domain auto-encoder architectures that use learned upsampling and downsampling layers, making the computational complexity of the model proportional to the output sampling rate. More recently, \ac{STFT}-based codecs\cite{funcodec}\cite{ncstft}\cite{apcodec}\cite{complex_dec} provide a more reliable way to work with different bandwidths without exploding the complexity of the model but still suffer difficulties at capturing finer details of high-frequency content of the signal.

Although neural codecs trained end-to-end offer promising solutions, the importance of \ac{BWE} cannot be overstated: It offers a modularity and scalability in terms of computational complexity, bitrate and signal quality that is attractive for many applications. A very similar approach to classical methods was proposed in~\cite{multimode} that used a neural codec for the \ac{WB} signal component and a classical transform-domain \ac{BWE} method to code higher bands. Neural networks have been widely explored for \ac{BWE} as well and have also commonly been referred to as audio super-resolution methods. These models either operate in the time domain~\cite{neuralvocoder}\cite{baenet}\cite{bweallyouneed} or in the frequency domain~\cite{aero}\cite{ATN_2020}\cite{apbwe} and are designed to either regenerate the entire signal or to only generate the missing frequency bands. These models are capable of performing high-quality \ac{BWE} but come at the cost of higher computational complexity and often inject additional algorithmic delay. Even the streaming capable models \cite{lc_streaming}\cite{LC_Wb2FB} are highly complex when considered for hand-held devices. Also, most research in this area is limited to expanding the bandwidth of non-coded speech signals.

Coded speech signals undergo severe degradation, especially at low bitrates. For best signal quality, these degradations need to be considered when generating high-frequency content from the low-frequency content. Some approaches~\cite{lpcganamr_nb}\cite{konstantin_amrnb}\cite{opusbwe} have been developed particularly for conventional speech codecs like \ac{AMR-NB} and Opus\cite{opus_codec}. However, enhancement of multiple speech codecs, including neural ones, with a single generalized \ac{BWE} model remains largely unexplored. Furthermore, to the best of our knowledge, fully end-to-end guided \ac{BWE} using neural networks has not yet been investigated. Although~\cite{schmidt2019deep} proposed a partially guided approach with a single bit of side information for fricatives, it falls short of full end-to-end guidance. In this paper, we present \ac{UBGAN}, a GAN-based \ac{BWE} model that is trained to work with multiple WB codecs to generate \ac{SWB} speech signals. Our contribution in this paper can be summarized as below:

%Furthermore, to the best of our knowledge, guided \ac{BWE} using neural networks has not yet been investigated. In this paper, we present \ac{UBGAN}, a GAN-based \ac{BWE} model that is trained to work with multiple WB codecs to generate \ac{SWB} speech signals. Our contribution in this paper can be summarized as below:

%Furthermore, fully end-to-end guided \ac{BWE} using neural networks has not yet been investigated. A partially guided approach using neural networks was explored in [23] where single bit was sent to indicate problematic frame, but the method is not end-to-end and significantly differs from our proposed system. In this paper, we present \ac{UBGAN}, a GAN-based \ac{BWE} model that is trained to work with multiple WB codecs to generate SWB speech signals. Our contribution in this paper can be summarized as below:

%Furthermore, fully end-to-end guided \ac{BWE} using neural networks has not yet been investigated. Though, a partially guided approach using neural networks was explored in~\cite{schmidt2019deep} by sending a single bit of side-information for problematic fricatives frames, but to the best of our knowledge, fully end-to-end guided \ac{BWE} using neural networks has not yet been investigated. In this paper, we present \ac{UBGAN}, a GAN-based \ac{BWE} model that is trained to work with multiple WB codecs to generate SWB speech signals. Our contribution in this paper can be summarized as below:
%

\begin{itemize}
	\item {} We propose a causal, lightweight \ac{BWE} model capable of extending \ac{WB} coded speech to \ac{SWB} for both conventional and neural speech codecs\footnote{Check our demo samples at: \url{https://fhgspco.github.io/ubgan/}}.
	\item {} We propose robust high band synthesis in the \ac{PQMF} subband domain, i.e., the proposed model is capable of generating high-frequency content even for a varying level of degradation in the low-frequency content. 
	\item {} We propose blind-\ac{UBGAN} that is a \ac{BWE} model conditioned on the mel-spectrogram calculated from the lower frequency bands of the coded speech signal.
	\item {} We also propose guided-\ac{UBGAN}, which relies on additionally transmitted quantized learned features from the upper \ac{PQMF} subbands for better steering of the \ac{BWE} system. The quantized features require an additional bitrate of 0.2~kbps.
\end{itemize}

%####################################Proposed model####################################
\section{Proposed Model}
\label{sec:Proposed}

\begin{figure}[t]
	\centering
    \includegraphics[width=0.90\columnwidth]{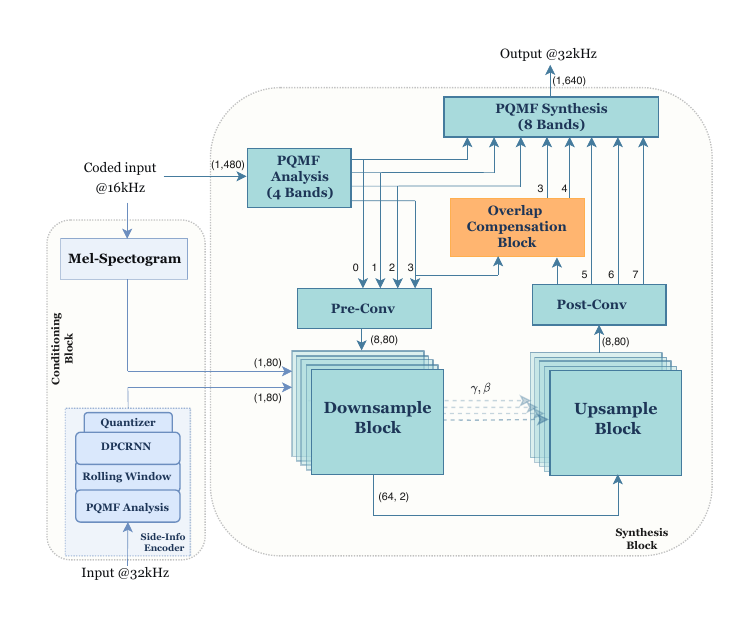}
	\vspace{-7pt}
    
	\caption{High-level schematics of \ac{UBGAN}. Dashed line represents residual connection.} %from downsample blocks to upsample blocks.}
    
 %The Mel-spectogram and side-info encoder block are part of the conditioning block, while the remaining components are part of the synthesis block and overlap compensation block.
	\label{fig:block_diagram}
\end{figure}

A schematic diagram of \ac{UBGAN} is shown in Fig.~\ref{fig:block_diagram}. The proposed model can be structured into three primary blocks, described in the following sections: %the conditioning block, the synthesis block, and an overlap-compensation block.

\subsection{Conditioning Block}

The conditioning block calculates features for conditioning the synthesis block. We use an 80-band Mel-spectrogram as conditioning for the blind-\ac{UBGAN} model. It is calculated from the coded signal on a frame size of 20~ms with 5~ms of past context and 5~ms of look-ahead to effectively capture temporal dependencies.

In contrast, the guided-\ac{UBGAN} model employs a neural side-info encoder block that is trained in an end-to-end manner alongside the synthesis block. This integrated approach enhances representation learning and synthesis quality. The side-info encoder consists of four key components: an eight-band analysis \ac{PQMF}, a rolling window mechanism, a \ac{DPCRNN} block~\cite{nesc} and a \ac{SQ} module~\cite{fsq}.  The input signal at sampling rate of 32~kHz, undergoes \ac{PQMF} analysis, resulting in eight subbands, each with 2~kHz bandwidth. To focus on higher frequency components, we discard the lowest four bands containing low-frequency signal components. The rolling window reshapes the data such that each frame consists of concatenated 20~ms segments from each of the higher \ac{PQMF} bands, accompanied by a 5~ms history and a 5~ms look-ahead. Each frame goes through a \ac{DPCRNN} block that consists of a \ac{GRU} sandwiched between two $1\times1$ convolution layers. The final output of the \ac{DPCRNN} is an 80-dimensional latent vector per frame which is then projected into a single dimension using a fully-connected layer before being quantized using 4~bits. Hence, \ac{GBWE} requires 0.2~kbps of additional bitrate.
 
%\vspace{-1pt}
\begin{figure}[t]
	\centering
	\vspace{-10pt}
    \includegraphics[width=0.90\columnwidth]{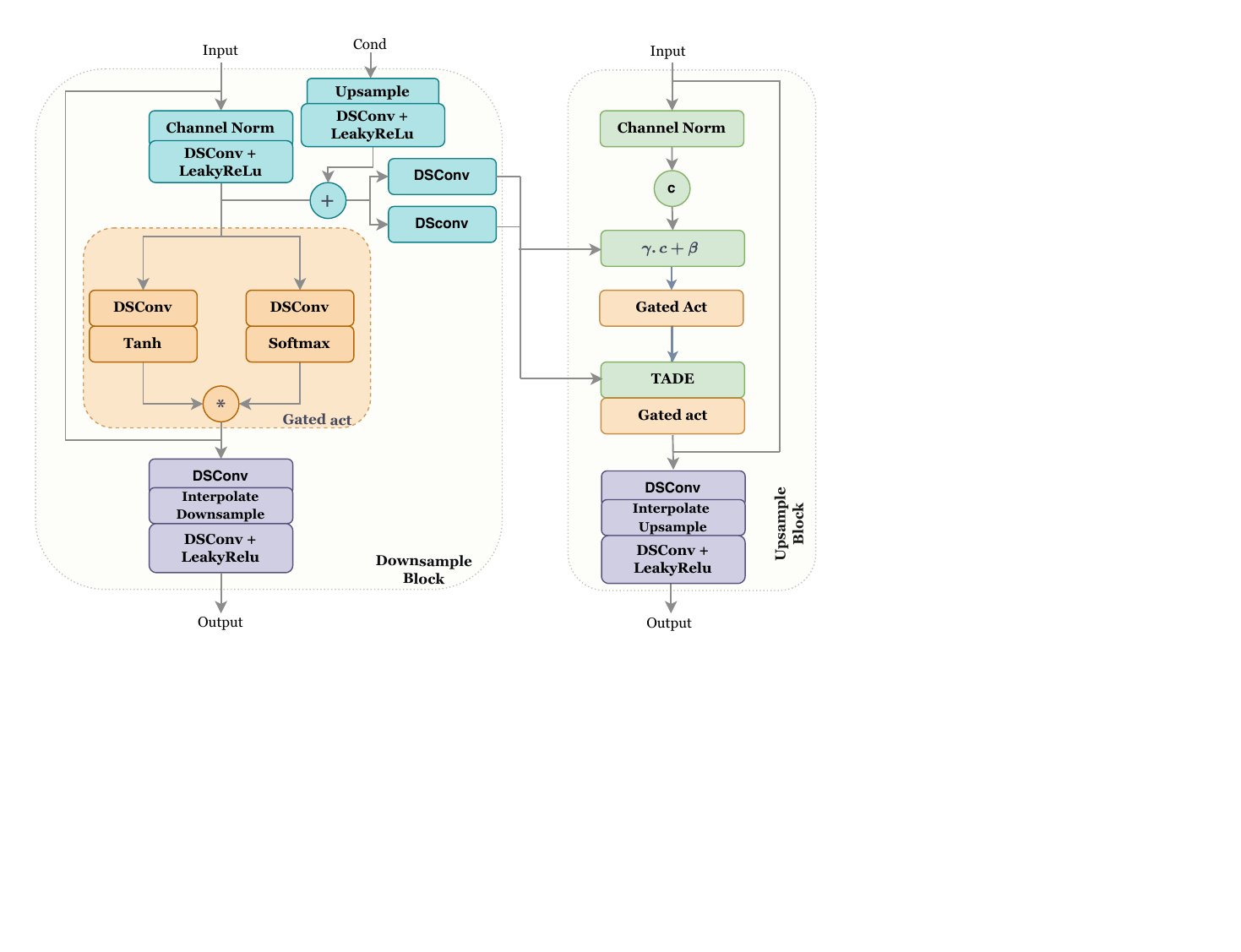}
	%\vspace{0.02\linewidth}
    \vspace{-10pt}
	\caption{Description of the downsample and upsample block of \ac{UBGAN}}
    \vspace{-9pt}
	\label{fig:ups_dwns_block}
\end{figure} 

\subsection{Synthesis Block}
The synthesis block is the pivotal part of the system. It works in the time domain and takes the coded signal along with the conditioning information as input. Initially, the \ac{WB} coded speech is  decomposed into four subbands using \ac{PQMF} analysis. The synthesis block then predicts additional four subbands, extending the bandwidth through an eight-band \ac{PQMF} synthesis. 

The architecture of the model is heavily inspired by PostGAN~\cite{postgan}, which comprises several key components, including \ac{PQMF} analysis and pre-conv, post-conv, downsample as well as upsample blocks. The pre and post-conv blocks contain a single convolutional layer with kernel size of 7 that maps the four input channels to eight output channels and vice-versa. The downsample and upsample blocks create a UNet-like structure with residual connections as shown in Fig.~\ref{fig:ups_dwns_block}.

In the downsample block, the input goes through channel normalization, 1-D convolution, gated activation and an interpolation layer sandwiched between two 1D convolutions~\cite{postgan}. We use six blocks with a varying number of channels [8, 16, 32, 48, 48, 64]. The downsampling of the input signal is done in each block via interpolation by a factor of [1, 2, 2, 2, 2.5, 2], respectively. The upsampling of the conditioning feature is done with scaling factors [80, 80, 40, 20, 10, 4] and the result is passed through a convolution layer. We obtain two 64-dimensional latent vectors for each 20~ms frame after the last downsample block. The output before the gated activation is merged with the upsampled conditioning signal to obtain the Temporal Adaptive DE-normalization~(TADE) parameters $\gamma$ and $\beta$ for the residual connection.
%%%bdl review till here
Each upsample block contains a TADE-residual block similar to~\cite{ssmgan}. The TADE parameters from the corresponding downsample block is used for performing an affine transformation of the normalized input. The operations normalization, linear transformation and gated activation are each applied twice before the output is upsampled. The sequences of channel numbers and upsampling scale factors are the reverse of the downsample block parameters. We use causal depthwise-separable convolutions~(DSConvs) in all the layers and a kernel size of 7. The use of DSConvs, along with a varying number of channels in the upsample and downsample blocks helps to achieve a model with low computational complexity and fewer parameters.

%, contributes to a model with low computational complexity and a lower parameter count.
%The varying number of channels in upsample and downsample block and use of DSConvs together helps to achieve model with low computaional complexity and fewer parameters. %Starting from PostGAN, the reduction in the number of used channels and the use of DSConvs enables a low complexity model for \ac{UBGAN}.

\subsection{Overlap Compensation Block}
In our \ac{PQMF} subband processing framework, the input coded speech is initially decomposed into four subbands, each with a bandwidth of 2~kHz. Subsequently, the output speech is synthesized using eight \ac{PQMF} subbands, with the four highest subbands being generated by the network. A simple concatenation of input and generated subbands produces audible artefacts due to overlap of the subbands and the inherent aliasing of the \ac{PQMF} decomposition. Specifically, the transition between the fourth and fifth subband requires extra effort to ensure that the aliasing between these two subbands cancels out effectively. To address this issue, we propose an overlap compensation block, which comprises two convolutional layers with tanh activation function. This block takes the highest subband of the coded speech and the lowest subband of the model-generated output, and outputs those subband signals with well-compensated overlap. This ensures a seamless transition during synthesis and eliminates potential artifacts.

 % at higher sampling rate.
%The network consist of 2D-convolution and takes concatenated real and imaginary part of STFT as input. 

%####################################Experimental Setup####################################
\section{Experimental Setup}
\label{sec:Experimental}

\subsection{Training and Losses}
We trained the models in a two-step procedure similar to~\cite{nesc}~\cite{ncstft}~\cite{postgan}. First, a pre-training is done for 50 epochs, then an adversarial training phase is started. Let $x$ and $\hat{x}$ denote the target and generated \ac{SWB} signals and $|X|$ and $|\hat{X}|$ their corresponding \ac{STFT} magnitudes. The pre-training loss consists of a spectral convergence loss
\begin{equation}\label{sc}
	\mathcal{L}_{\text{sc}} = \frac{\left\| \, |X| - |\hat{X}| \, \right\|_F}{ \left\| \, |X| \, \right\|_F}
\end{equation}
and a log-magnitude loss
\begin{equation}\label{mag}
	\mathcal{L}_{\text{mag}} = \frac{1}{N} \left\| \, \log(|X|) - \log(|\hat{X}|) \, \right\|_1,
\end{equation}

where $\left\| \,  . \, \right\|_F$, $\left\| \, . \, \right\|_1$ and $N$ represents the Frobenius and the L1 norms and the number of time-frequency bins, respectively. As adversarial loss for training the generator we use the Least-Squares GAN~(LSGAN) loss~\cite{lsgan}
\begin{equation}\label{adv}
	\mathcal{L}_{\text{adv}} = \mathbb{E}_{p(x)} \left[ \sum_{k=1}^{4} \left( D_k(G(x)) - 1 \right)^2 \right],
\end{equation}

where $D_k$ represents the $k^{th}$ discriminator for generated signal $G(x)$, and $p(x)$ denote the probability density of the training data $x$. Alongside the LSGAN loss, we use a feature loss that calculates the L1-loss between features from the generated signal and the ground truth signal at the output of each discriminator layer. Finally, the feature loss is scaled by a factor of 10 while all other losses are added unscaled.

For generator training, we use the ADAMW optimizer with learning rate $\text{lr}_\text{g} = 5e\text{-}4$ and exponential decay by a factor of 0.99 every five epochs. The discriminator is trained with the ADAM optimizer with learning rate $\text{lr}_\text{d} = 2e\text{-}4$. We use a batch size of 64 and the straight-through estimator is used at the quantizer bottleneck for training the side-info encoder of guided-\ac{UBGAN}. 

\subsection{Discriminator}
For adversarial training, we use an ensemble of four \ac{STFT} discriminators proposed in~\cite{encodec}. The discriminators of the ensemble are working on multiple window lengths (2048, 1024, 512 and 256) with 75\% overlap. We also tried the multi-scale discriminator of \cite{ncstft}\cite{nesc} but found that such time-domain discriminators do not perform as well as their frequency-domain counterpart for \ac{SWB} signals.

\subsection{Dataset}
The model was trained with 562 hours of the read speech dataset~\cite{read_speech} and 44 hours of the VCTK dataset~\cite{VCTK}. Both datasets contain \ac{FB} signals which are resampled to 16~kHz for input and 32~kHz for SWB output signals. The model is trained with coded speech from two conventional codecs and two neural codecs. We encode and decode each 16~kHz signal using the following codecs with equal probability: EVS 5.9~kbps~\cite{evs_overview}, Opus 6~kbps~\cite{opus_codec}, LyraV2 3.2~kbps~\cite{soundstream}, STFTNC 1.5~kbps~\cite{ncstft}. For both subjective and objective evaluation, we select a diverse set of 24 items created from various sources including ITU-T Standard P.501~\cite{ITU}, ETSI TS 103 281 Annex E~\cite{ETSI-TS} and NTT~\cite{nttdb:2012}.  

\subsection{Evaluation}
For evaluation, we carry out both subjective and objective evaluations. For objective assessment, we use the 2f-Model~\cite{2f_model} and the VISQOLAudio V3 metric~\cite{visqol} as they both support evaluation over FB speech signals. For subjective assessment, we conduct a P.808 Degradation Category Ratings (DCR) listening test~\cite{p808} using the Amazon Mechanical Turk service.

\subsection{Baselines}
In our evaluation, we employed a diverse set of codecs at various bitrates, which we classify in the following categories:

\begin{itemize}[noitemsep, leftmargin=*]
	\item \textbf{Seen WB codecs:} LyraV2~(3.2~kbps)~\cite{soundstream}, STFTNC~(1.5~kbps)~\cite{ncstft}, EVS~(7.2~kbps)~\cite{evs_overview}, Opus~(9~kbps)~\cite{opus_codec}. Note that the bitrates used for EVS and Opus are different than in training.
	\item  \textbf{Unseen WB codecs}:  Funcodec~(3~kbps)~\cite{funcodec}, LPCNet~(1.6~kbps)~\cite{lpcnet}, AMR-WB~(6.6~kbps)~\cite{amrwb}.
	\item  \textbf{Semi-SWB/SWB codecs}: DAC-IBM~(1.5~kbps)~\cite{dac_ibm}, Encodec~(3~kbps)~\cite{encodec}, EVS-SWB~(9.6~kbps)~\cite{evs_swb}. The outputs of these models were resampled to 16~kHz before \ac{BWE}.
	\item  \textbf{BWE-model}: We use the STFT-based AP-BWE~\cite{apbwe} model as baseline. It performs \ac{FB} extension of the \ac{WB} speech up to a bandwidth of 48~kHz.
\end{itemize}

%####################################Results####################################
\section{Results and Discussion}
\label{sec:Results}

\begin{table}[t]
	\centering
	\vspace{1.5pt}
	\resizebox{\columnwidth}{!}{
	\begin{tabular}{l*{4}{S[round-precision=2,table-format=2.1]S}}
		\toprule
		& \multicolumn{4}{c}{\textbf{2f-Model}$\uparrow$} & \multicolumn{4}{c}{\textbf{VISQOLAudio}$\uparrow$}\\
		\cmidrule(lr){2-5}\cmidrule(lr){6-9}
		& \textbf{WB}& \textbf{Blind} & \textbf{Guided} & \textbf{AP-BWE} & {WB}& \textbf{Blind} & \textbf{Guided} & \textbf{AP-BWE}\\
		\midrule
        \addlinespace[0.15cm]
        \vspace{0.15cm}
		AMR-WB~(6.6~kbps)   & 19.31  & 31.55  & \textbf{34.64} & 22.87  & 2.79  &  3.33  &  \textbf{3.54} & 2.64\\
        \vspace{0.15cm}
		EVS~(7.2~kbps)      & 20.85  & 31.50  & \textbf{36.26} & 25.30  & 2.86  &  3.28  &  \textbf{3.59} & 2.88  \\
		Opus~(9~kbps)       & 28.64  & 39.48  & \textbf{41.40} & 35.21  & 2.95  &  3.56  &  \textbf{3.64} & 3.02  \\
		\midrule
        \addlinespace[0.15cm]
        \vspace{0.15cm}
		FunCodec~(3~kbps)   & 24.28  & 35.06  & \textbf{38.52} & 29.78  & 2.88  &  3.42  &  \textbf{3.55} & 2.94  \\
		\vspace{0.15cm}
		LPCNet~(1.6~kbps)   & 23.20  & 32.29  & \textbf{33.49} & 27.67  & 2.72  &  3.05  &  \textbf{3.36} & 2.87  \\
		\vspace{0.15cm}
		LyraV2~(3.2~kbps)   & 25.42  & 36.21  & \textbf{37.51} & 31.59  & 2.86 &  3.48  &  \textbf{3.53} & 3.07  \\

		STFTNC~(1.5~kbps)   & 23.06  & 31.63  & \textbf{34.38} & 27.54  & 2.84  &  3.36  &  \textbf{3.50} & 2.97  \\
		\bottomrule
	\end{tabular}}
	\vspace{0.8pt}
	\caption{Average VISQOL-Audio and 2f-Model scores of our proposed method and the baseline. Blind and guided denote the resulting SWB signals from blind- and guided-UBGAN, respectively. AP-BWE is the baseline BWE method for obtaining FB speech. For all conditions, the FB signals are used as reference.} 
    \label{tab:obj_score}
    \vspace{-5pt}
\end{table}

In Table~\ref{tab:obj_score}, we present the objective scores obtained for various codecs with our proposed and baseline \ac{BWE} methods. Both objective measures are reference-based, thus for uniformity, for each measure, the FB signals are used as reference irrespective of the bandwidth of the signals under test. The 2f-Model as well as the VISQOL scores show the quality improvement obtained through BWE over \ac{WB} speech. For blind- and guided-UBGAN, both metrics show the superiority of the guided model which is able to reconstruct higher frequencies better. Despite higher bandwidth, the AP-BWE method scores lower than both of our proposed methods. Upon closer inspection, we found that the baseline method AP-BWE that was trained on clean speech injects the coding artefacts from the lower bands to higher bands, which degrades the quality of the resulting \ac{FB} signal.

\begin{figure}[t]
	\centering
	\includegraphics[width=0.8\columnwidth]{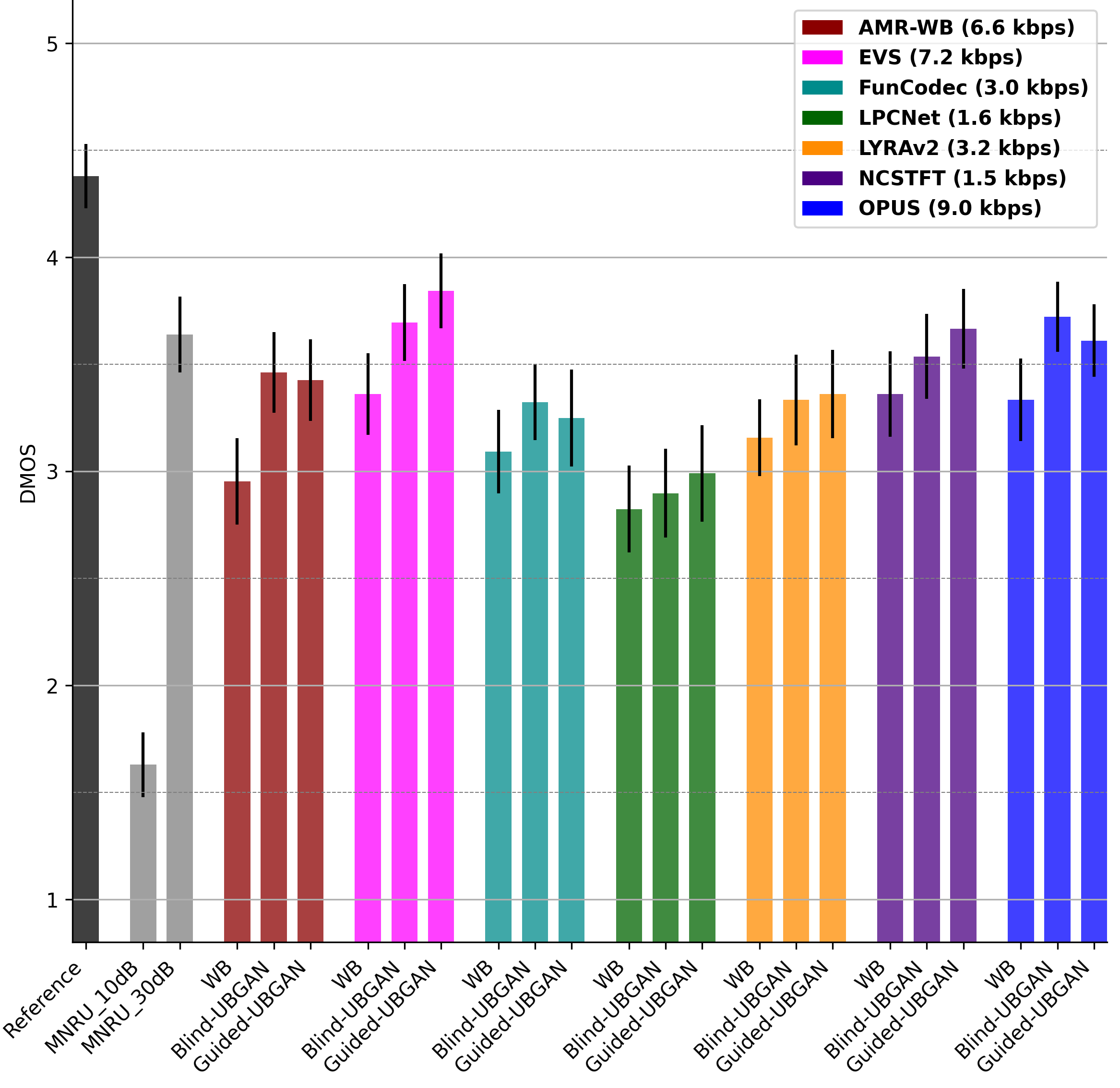}
    \vspace{-11pt}
	\caption{P.808 DCR scores with 27 listeners for WB codecs with blind-\ac{UBGAN} and guided-\ac{UBGAN}.}
	\label{fig:subj_score_v1}
    \vspace{-10pt}
\end{figure}

For a better assessment of our proposed method, we conducted two P.808 DCR subjective listening tests. DCR tests include the presentation of the reference signal to the listeners before the degraded items are presented. In the first test, we evaluated the benefit of our proposed \ac{BWE} method over WB codecs. The results shown in Fig.~\ref{fig:subj_score_v1} clearly indicate the improvement in quality brought by wider bandwidth. The guided-UBGAN performs slightly better than the blind variant for all codecs except for Funcodec and OPUS. We investigated the individual scores of the items for these codecs and found that, most items in the guided case are rated almost on par with or above the blind-UBGAN except for a few stimuli where degradation of the coded signals seems to be substantial. Upon casual listening, we established that the core-codec quality has a major impact on the perceived quality of the extended bandwidth. Here, the guided \ac{BWE} version can lead to a mismatch between the energy levels of the extended bandwidth components and the coded core bandwidth components. This is because this model is conditioned by the encoded high frequency side-information derived from the clean signal, which might be of different energy than the coded signals. In contrast, the blind-UBGAN is more consistent since it relies solely on the coded signal and adapts to its distortions.

%In contrast, the blind-UBGAN extrapolates the energy and fine structure from coded bandwidth.

\begin{figure}[t]
	\centering
	\includegraphics[width=0.8\columnwidth]{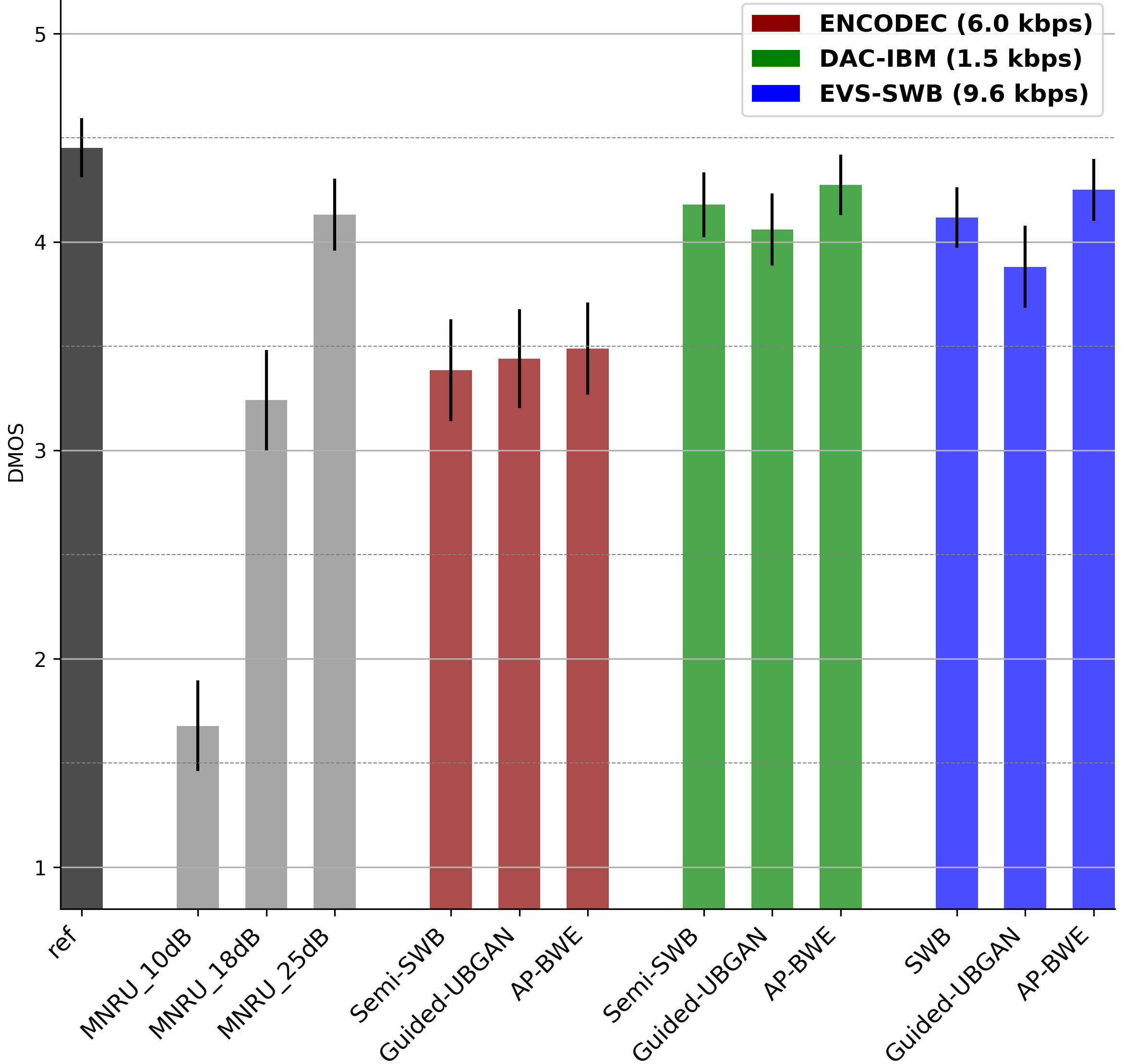}
    \vspace{-10pt}
	\caption{P.808 DCR scores with 21 listeners for semi-SWB/SWB codecs with guided-\ac{UBGAN} and AP-BWE baseline}
	\label{fig:subj_score_v2}
    \vspace{-10pt}
\end{figure} 

The results shown in Fig.~\ref{fig:subj_score_v2} compare semi-SWB/SWB codecs, our proposed guided-\ac{UBGAN}, and the output of the pre-trained AP-BWE model capable of full-band extension of the signal. The core-coded signals are resampled to 16~kHz before \ac{BWE} and the used reference signal has a sampling rate of 48~kHz. For the neural codecs with 12~kHz bandwidth, our proposed approach that reconstructs the lost bandwidth due to resampling, performs similar to the semi-SWB baseline codecs themselves, thus showing the effectiveness of our modular approach compared to the end-to-end approaches. However, for the conventional SWB codec, EVS, the proposed approach falls short. This may be attributed to the fact that the parametric-\ac{BWE} in EVS uses a higher bitrate of 0.95~kbps to perform guided \ac{BWE} and is highly tuned to the core-coder and the given operating point. Also, our proposed model was trained on lower bitrates and lower signal qualities of core-codecs and, thus, has limitations for codecs achieving higher signal quality. For almost all codecs, the computationally complex AP-BWE method improves signal quality relative to the codec output and the proposed solution, but the perceived quality difference is not substantial. From our results, we also observe that the FB extension of speech signal might not lead to substantial quality improvements and semi-SWB/SWB or WB codecs associated with (semi) SWB-\ac{BWE} can be seen as efficient design choices, especially in the paradigm of neural codecs, where model complexity increases with the core-bandwidth.

\begin{table}[t]
	\begin{center}
			\resizebox{0.90\columnwidth}{!}{
			\begin{tabular}{c  c  c }
				\toprule
				\textbf{Models} & \textbf{Complexity (GFLOPS)} & \textbf{Parameter Count}  \\ 
				\midrule
				AP-BWE (16~kHz to 48~kHz)   & 17.87  &  29.76M  \\
				Blind-\ac{UBGAN}   & 0.214  &  250K   \\
				Guided-\ac{UBGAN}   & 0.246 &  364K    \\
				\bottomrule
			\end{tabular}}
            \vspace*{2pt}
		\caption{Complexity and parameter count of the proposed model and baseline}
		\label{tab:complexity} 
		\vspace{-15pt}
	\end{center}
\end{table}

The computational complexity and the parameter count of the compared models are shown in Table~\ref{tab:complexity}. Our proposed blind- and guided-\ac{UBGAN} models are significantly less complex than the AP-BWE and also have fewer parameters. The resulting computational efficiency enables the model to run on devices with low computational power in real time and also facilitates their integration with other neural codecs while maintaining minimal complexity overhead. Furthermore, the proposed models are causal and operate on a frame size of 20~ms and a look-ahead of 5~ms, making them suitable for integration into a variety of codecs without introducing significant algorithmic delay. 

%####################################Conclusion####################################
\section{Conclusion}
\label{sec:conclusions}
In this paper, we introduced a universal low-complex, low-delay and causal \ac{BWE} method for \ac{WB} speech codecs. We demonstrate its effectiveness across a wide range of speech codecs. The proposed model works in the \ac{PQMF} domain and synthesizes the higher subbands of the signal. Furthermore, we introduced a guided version of it that learns and quantize high-frequency features at a low bitrate of 0.2~kbps from the original speech. Our findings indicate that the guided-\ac{UBGAN} model offers significantly greater quality improvements than the blind-\ac{UBGAN} for most codecs. Additionally, our analysis highlights the importance of bandwidth in perceived speech quality. It also emphasizes that for low-bitrate codecs, the quality of the speech signal tends to saturates at semi-SWB or SWB and the advantage of performing \ac{FB} extension might be limited. In future work, we intend to extend the model to general audio applications and investigate ways to improve the performance of guided methods. 

% remove in double blind review
\section{Acknowledgements}
\label{sec:Acknowledgements}
We would like to thank Markus Multrus and Anika Treffehn from Farunhofer IIS for their support. Parts of this work have been funded by the Free State of Bavaria in the DSgenAI project. We thankfully acknowledge the scientific support and HPC resources provided by the Erlangen National High Performance Computing Center (NHR@FAU).

\clearpage

\bibliographystyle{IEEEtran}
\bibliography{refs25}

\end{document}